\documentclass[%
twocolumn,
 amsmath,amssymb,
 aps,
prb,
longbibliography,
]{revtex4-2}

\usepackage{hyperref}
\hypersetup{colorlinks,linkcolor={blue},citecolor={blue},urlcolor={blue}}  

\usepackage{xcolor} 

\usepackage{graphicx}
\usepackage{dcolumn}
\usepackage{bm}
\usepackage{xcolor}
\usepackage{mathtools}
\usepackage{extpfeil}

\newtheorem{proposition}{Proposition}
\newtheorem{mydef}{Definition}
\newtheorem{remark}{Remark}

\begin{document}

\preprint{APS/123-QED}

\title{Heat and Work in Quantum Thermodynamics:\\a Cybernetic Approach}

\author{William Rupush}
 \email{william.rupush@physics.uu.se}
 \affiliation{Department of Physics and Astronomy, Uppsala University}
\author{Oscar Grånäs}%
 \email{oscar.granas@physics.uu.se}
\affiliation{Department of Physics and Astronomy, Uppsala University}

\date{\today}

\begin{abstract}
We present a new proposal for distinguishing heat from work based on a control-theoretic observability decomposition. We derive a Hermitian operator representing instantaneous dissipation of observable energy, and suggest a generalization of the von-Neumann entropy which can account for the model-uncertainty also present in pure states if the measured observables are informationally incomplete. In this view, the transition from a fundamental to a thermodynamic model consists in mapping the fundamental density matrix to an effective one, generally of lower dimension, encoding only what is observable given the constraints of our sensor and actuator capabilities. The generalized entropy captures the information loss incurred in this mapping. The theory is illustrated for the central spin model, where we show that the application of external controls can increase the size of thermal fluctuations and lower the entropy.
\end{abstract}

\maketitle


\section{\label{sec:level1}Introduction: Heat and Work\protect\\ }

The First Law of Thermodynamics, aside from being a statement of energy conservation, claims that there are two qualitatively different ways to alter the energy of a thermal system: by the application (or extraction) of \emph{work}, or the addition (or subtraction) of \emph{heat}. How to make this distinction precise is a major conceptual issue in thermodynamics, with fundamental ramifications. A satisfactory solution is still lacking, and this presents a roadblock towards the development of a mature science of quantum thermodynamics \cite{Alicki2018IntroductionProspectsb,2018ThermodynamicsRegime}. \\ \indent
A common approach is to proceed in a heuristic manner and make sure that the problem under consideration contains a subsystem which, whatever the exact definitions of heat and work, is one which would surely be recognized as either a heat or a work reservoir, or in other words, as a bath or a battery \cite{Mahler2014QuantumProcesses}. The bath is usually taken to be an infinite collection of quantum harmonic oscillators, and the battery as any system whose energy can be read off from measurements, and efficiently stored and extracted on demand. In this way the issue of precise definitions is sidelined for particular applications. Alternatively, in order to obtain a heat-work decomposition for interactions with a more general environment, one may follow R.Alicki \cite{Alicki1979TheEngine} and consider the following application of the Leibniz product rule
\begin{align}
\frac{d}{dt} \langle \hat H \rangle = \langle \frac{d}{dt} \hat H , \hat \rho \rangle + \langle \hat H , \frac{d}{dt} \hat \rho \rangle,
\end{align}
and decide that the first term corresponds to "work", and the second to "heat". This method is applied widely in the research literature to reduced density matrices in open systems \cite{Ahmadi2023OnSystems,Vallejo2021QubitRegime}. It can be shown that under certain constraints imposed on the Hamiltonian, only the second term contributes to an increase in the local von-Neumann entropy. A slightly more general method which relaxes said constraints is obtained using a so called \emph{local effective measurement basis} (LEMBAS) which is determined by the experimental set-up \cite{Hossein-Nejad2015WorkSystems}. In either of the aforementioned views, only open quantum systems can thermalize, and any heat-work decomposition requires a partitioning of the total system into a "working medium" of interest, and an "external bath". But recent experiments \cite{Kaufman2016QuantumSystem,Langen2015ExperimentalEnsemble,Gring2012RelaxationSystem,Trotzky2012ProbingBosegas} have indicated that closed systems thermalize as well, stimulating new interest in alternative methods, such as the eigenstate thermalization hypothesis \cite{Deutsch2018EigenstateHypothesis}. In the view proposed in this paper, both closed and open systems can thermalize, and instead of proclaiming a system-bath split by fiat it is derived based on the measurement and control resources at hand.  \\ \indent
The two central notions of control theory are \emph{controllability} and \emph{observability}. The first deals with the extent to which state transformations can be actuated by varying externally controllable parameters, and the second the extent to which the state of the system can be inferred from measurements. Control-theorists are interested in separating out degrees of freedom which are controllable and/or observable, from those which are not \cite{Sontag1998MathematicalTheory}. We claim that it is only when the system is not fully observable, that is, when at least a part of it is a ''black box'', that thermodynamics as typically conceived is applicable. "Entropy" appears when degrees of freedom are hidden from us. This is effectively the case in classical thermodynamics where we only have access to course-grained macroscopic observables, and fundamentally so for an event horizon surrounding a black hole, obscuring the interior for outside observers. In our view, the transition from a fundamental to a thermodynamical model amounts to a model-reduction procedure where the unobservable degrees of freedom have been discarded, here described as a mapping to an \emph{effective density matrix}. Using this mapping we obtain an entropy function that is non-vanishing also for pure states, provided at least a part of the system is unobservable, and is in general time-varying for an isolated system. \\ \indent
We claim that the control-theoretic notion of observability is needed for a proper understanding the heat-work decomposition: \emph{heat is energy flow into unobservable degrees of freedom, and work into those which are observable}. Or in other words, \emph{work is energy transfer into information-bearing degrees of freedom}. While the use of the observability decomposition to this end is novel the general idea is to some extent already familiar to physicists, who have recognized the connection between information and thermodynamics ever since Maxwell formulated his famous thought experiment with the demon \cite{Maxwell2011TheoryHeat}, and even more so since Shannon introduced his information-theoretic entropy \cite{Shannon1948ACommunication}, and Landauer's discovery of the link between heat production and information erasure \cite{Landauer1996TheInformation}. Our theory solidifies this connection and brings together the fields of quantum control, information and thermodynamics in a very explicit manner.
\\ \indent 
We stress that the concepts of heat, work and entropy presented here are not universal, but explicitly context-dependent and specific for any given quantum control system. The heat-work decomposition cannot be performed for a dynamical system without specifying our means for controlling and observing it. This philosophy claims that thermodynamics is as much a theory of engineering as of physics per se. This should not be surprising to the historian of science who is aware of the fact that thermodynamics was developed during the industrial revolution with the expressed purpose of optimizing the performance of heat engines. \\ \indent
This formalism by itself does not force the usual thermodynamical laws to hold; this would require assumptions about the actual physics of the system under study. There is no universal law stating that energy must tend to flow into unobservable degreees of freedom (i.e. for entropy to increase), but this becomes highly probable if we assume that the dimensionality of the observable degrees of freedom to be much smaller than that of the unobservable, as well as sufficient coupling between the two. Its value therefore lies in clarifying fundamental concepts, rather than replacing the other approaches. For example, statistical mechanics is still required to understand the route towards and conditions for equilibrium. Paradigmatic applications we have in mind include the calculation of the energy dissipated in quantum computation processes, and the construction of impurity models. \\ \indent
After giving an overview of the relevant concepts from the quantum control theory literature, where D. D'Alessandro's observability decomposition \cite{DAlessandro2003OnMeasurement} is at center stage, we formulate our proposal explicitly and illustrate its application in the central spin model, which is experimentally relevant as it can be used to model decoherence in NV centers \cite{Park2022DecoherenceDiamond} and quantum dots \cite{Bechtold2015Three-stageDot,Arfaoui2022DecoherenceDot}. Along the way we also prove a proposition connecting the observability decomposition to a measure of Fisher information.

\section{\label{sec:level2}Quantum control theory \protect\\ }
An up to date overview of the current state of quantum control theory can be found in \cite{DAlessandro2021IntroductionDynamics}. Consider a quantum system in the density matrix formalism governed by the Liouville-von-Neumann equation
\begin{align}
\label{lvne}
\frac{d}{dt} \hat \rho = - i [\hat H (u) , \hat \rho],
\end{align}
where the Hamiltonian depends on a function $u$ freely selected by the control engineer from a set of admissible controls $\mathbb{U}$, generally assumed to be piece-wise constant. The Hamiltonian is assumed to be of bilinear form
\begin{align}
\hat H (t) = \hat H_0 + \sum_{j=1}^k u_i (t) \hat H_i,
\end{align} 
where $\hat H_0$ is referred to as the \emph{drift Hamiltonian}, and $\hat H_i$ for $i \in \{ 1,...,k \}$ as \emph{control Hamiltonians}. The measured output is the expectation value of a Hermitian operator $\hat S$, i.e. $y(t) = \langle \hat S , \hat \rho \rangle$. If $\mathcal{H}$ is the Hilbert space of states under consideration, we will refer to the quadruple $\Sigma = (\mathcal{H}, \hat H( \cdot ), \mathbb{U}, \hat S)$ as a \emph{quantum control system}. Let $n$ denote $\text{dim} \mathcal{H}$.
\subsection{Controllability}
The \emph{operator controllability problem} consists of determining the \emph{reachable set}
\begin{align*}
\mathcal{R} \triangleq \big \{ \hat U \in U(n) \text{ } \lvert \text{ } \hat U = \hat U_u (t) \text{ for some } t \geq 0 , u \in \mathbb{U} \big \},
\end{align*}
where $\hat U_u$ satisfies the operator Schrödinger equation
\begin{align}
i \frac{\partial}{\partial t} \hat U_u = \hat H(u) \hat U_u.
\end{align}
$\mathcal{R}$ is the set of all unitary transformations which can be achieved by varying $u$ over $\mathbb{U}$. If $\mathcal{R} = U(n)$ then any unitary matrix can be obtained by a suitable choice of controls $u_1 , ... , u_k$ and time $t$, in which case the system is said to be \emph{operator controllable}, the strongest of the controllability notions considered in the literature. \cite{Albertini2003NotionsSystems} \\ \indent 
A central theorem of quantum control theory states that the reachable set can be obtained as the exponential image of the \emph{dynamical Lie algebra} \cite{Huang1983OnSystems}, i.e.
\begin{align}
\mathcal{R} = e^{\mathcal{L}},
\end{align}
where $\mathcal{L}$ is the Lie algebra generated by the set $\{-i \hat H_0 , -i \hat H_1 , ... , - i \hat H_k \}$. It is the vector space of all real linear combinations of elements in this set, as well as of their nested commutators. It follows from the above that the system is operator controllable if $\mathcal{L} = \textbf{u}(n)$ or $\mathcal{L} = \textbf{su}(n)$. The difference between the two cases amounts to whether one can control the overall phase of the system, a difference generally considered to lack physical significance.

\subsection{Observability}
A quantum control system is said to be \emph{observable} if its internal state can be determined from measurements of the chosen observable(s). Observability is defined formally with reference to the  equivalence relation of \emph{indistringuishability}. Let $ \hat \rho (t,u, \hat \rho_0)$ denote the solution to Eq. \ref{lvne} for the initial state $\hat \rho_0$ and applied control $u$, and assume that we are measuring an observable $\hat S$.
\begin{mydef}
\label{defin}
A pair of states $(\hat \rho_0 , \hat \rho_0')$ are indistinguishable, denoted by $\hat \rho_0 \sim \hat \rho_0'$, if for any control $u \in \mathbb{U}$ we have
\begin{align}
\langle \hat S, \hat \rho (t,u, \hat \rho_0) \rangle = \langle \hat S, \hat \rho (t,u, \hat \rho_0') \rangle, \indent \forall t \geq 0.
\end{align}
The quantum control system $\Sigma$ is observable in one step if
\begin{align}
\hat \rho_0 \sim \hat \rho_0' \iff \hat \rho_0 = \hat \rho_0'.
\end{align}
\end{mydef}
This definition has been generalized to both multiple and non-projective measurements \cite{DAlessandro2006FurtherMeasurement}, but here we assume a single projective measurement. A necessary and sufficient condition for observability (in one step), due to D. D'Alessandro \cite{DAlessandro2003OnMeasurement}, makes use of an object called the \emph{observability space}, defined
\begin{align}
\mathcal{V} \triangleq \bigoplus_{j=0}^{\infty} \text{ad}_{\mathcal{L}}^j \text{span} \{ i \hat S' \},
\end{align}
where $\hat S' = \hat S - \frac{\text{Tr} (\hat S)}{n} \hat I$. If the measured observable $\hat S$ is an element of $i \mathcal{L}$, then $\mathcal{V}$ is the ideal of $\mathcal{L}$ generated by $\hat S$. A quantum control system is observable in one step if an only if
\begin{align}
\mathcal{V} = \textbf{su}(n).
\end{align}
Note that if $\mathcal{L} = \textbf{su}(n)$, then since $\textbf{su}(n)$ is simple (i.e. contains no non-trivial ideals), $\Sigma$ is observable for any $\hat S$ not proportional to the identity operator.
If the system is not observable, then there exists a decomposition $\hat \rho = \hat \rho_o + \hat \rho_u$, where $\hat \rho_o \in i \mathcal{V}$ and $\hat \rho_u \in i \mathcal{V}^{\perp}$, such that the measured output depends only on $\hat \rho_o$:
\begin{align}
\label{output}
y(t) = \frac{1}{n} \text{Tr} (\hat S) + \langle \hat S , \hat \rho_o \rangle.
\end{align}
The orthogonal complement $\mathcal{V}^{\perp}$ is taken inside of $\textbf{su}(n)$ with respect to the Hilbert-Schmidt inner product. If $\hat S \in  i \mathcal{L}$ then the reachable set splits into a semi-direct product of Lie groups
\begin{align}
e^{\mathcal{L}} \simeq e^{\mathcal{V}} \rtimes e^{\mathcal{L}/\mathcal{V}}.
\end{align}
This yields a dynamical decomposition where the action of quotient group leaves the observed output unaffected. The condition $\hat S \in i \mathcal{L}$ is obtained in all cases where one is measuring a local observable $\hat S\otimes \hat I$, on a locally controllable system, meaning a system where $\textbf{su}(\tilde N) \otimes \hat I \subset \mathcal{L}$ and $\tilde N = \text{dim} (\hat S)$. \\ \indent
To obtain the observability decomposition we project the density matrix onto each of the basis elements of $i \mathcal{V}$. Provided the basis is orthogonal, we have
\begin{align}
\hat \rho_o (t) = - \sum_{i=1}^{\text{dim} \mathcal{V}} \langle \hat B_i, \hat \rho (t)\rangle \frac{\hat B_i}{\lvert \lvert \hat B_i \rvert \rvert^2}.
\end{align}
$\mathcal{V}$ is not itself necessarily a Lie algebra, but when it is, which includes the important case when $\hat S \in i \mathcal{L}$, work can be distinguished into qualitatively different kinds using any of the known Lie algebra decompositions. Notably, using the Levi decomposition \cite{DAlessandro2021IntroductionDynamics} we can split $\mathcal{V}$ into a direct sum $\mathcal{Z} \otimes \mathcal{S}$, where $\mathcal{Z}$ is abelian and $\mathcal{S}$ semi-simple. $\mathcal{V}$ then assumes the form of a set of parallel subsystems that are mutually commuting and orthogonal, plus a classical subsystem $\mathcal{Z}$ consisting of only commuting observables. In this way we can differentiate "classical" from "quantum" work depending on whether the energy goes into the classical system $\mathcal{Z}$, or any of the simple quantum systems in $\mathcal{S}$.
\begin{remark}
\label{rem}
The key property which characterizes isolated thermal states is that observables have expectation values proportional to the trace of the corresponding Hermitian operator, i.e. $y(t) \propto \text{Tr} (\hat S)$. This is referred to as the microcanonical ensemble. Note that this condition, for a specified observable $\hat S$, is equivalent to the condition that $\hat \rho \in i \mathcal{V}^{\perp}$ by Eq. \ref{output}. This implies that the maximally thermalized state is the maximally unobservable one, and therefore reveals a new framework for thinking about thermalization.
\end{remark}
\subsection{Observability and Fisher Information}
The \emph{Fisher information} is a key concept in both classical and quantum estimation theory which represents the extent to which measurements on random variables provides information on some unknown parameter of interest \cite{Liu2020QuantumEstimation,Yu2022QuantumQubit}. The Fisher information can be defined in a variety of ways, but here we'll focus on a particular kind defined using \emph{symmetric logarithmic derivative (SLD) operators} \cite{Liu2016QuantumAnti-commutators}. \\ \indent
Consider a density matrix $\hat \rho (\theta)$ dependent on $r$ parameters $(\theta_1, ... , \theta_r)$. The matrix components of the Fisher information are then defined by the equation
\begin{align}
F_{ij} \triangleq \frac{1}{2} \langle \hat \rho, [\hat L_{\theta_i} , \hat L_{\theta_j}]_+ \rangle,
\end{align}
where the SLD operators $\hat L_i$, for $i \in \{ 1 , ... , r \}$ are defined by the equations
\begin{align}
\frac{\partial}{\partial \theta_i} \hat \rho (\theta) = \frac{1}{2} [\hat L_{\theta_i} , \hat \rho]_{+}.
\end{align}
A key property of the Fisher information is that it provides a bound on covariances of unbiased estimators $\tilde \theta$ of $\theta$, a result called the Cramer-Rao inequality:
\begin{align}
\text{Cov} (\tilde \theta) \geq F^{-1}.
\end{align}
The Fisher information is typically a complicated time-dependent function, but when the parameter $\theta$ is a component of the state along a basis vector $\hat{B_j}$ of $i \mathcal{V}$ it assumes a particularly simple form. $\hat{B_j}$ is then an unbiased estimator of $\theta_j$ for every $j \in \{ 1,..., \text{dim} \mathcal{V} \}$.
\begin{proposition}
\label{fishprop}
Let $\Sigma$ be a quantum control system with $\{ \hat B_i \}$ the basis of its observability space, and let the estimated parameters be the projections $\theta_i = \langle \hat \rho, \hat B_i \rangle$. If the quantum state is pure, then the SLD Fisher information is the metric tensor of the observability space induced by the Hilbert-Schmidt inner-product:
\begin{align}
F_{ij} = \langle \hat B_i , \hat B_j \rangle.
\end{align}
\end{proposition}
A proof is found in the appendix.
\section{Quantum dissipation channels}
We define a \emph{quantum dissipation channel} to be any component $ i\hat H_u \in \mathcal{V}^{\perp}$, of an element $i \hat H \in \mathcal{L}$, such that $[\hat H, \hat H_u] \neq 0$. In the particular case where $i \hat S \in \mathcal{L}$, when $\mathcal{V}$ is an ideal of $\mathcal{L}$, we have the short exact sequence
\begin{align}
\mathcal{V} \xhookrightarrow[]{i} \mathcal{L} \xtwoheadrightarrow{\pi} \mathcal{L}/\mathcal{V} ,
\end{align}
where $i$ is the inclusion map and $\pi$ the canonical projection. Then a set of dissipation channels can be viewed as a non-abelian Lie algebra extension of $\mathcal{V}$ in $\mathcal{L}$. In this case a quantum system has a dissipation channel if the observability space is embedded in a strictly larger dynamical Lie algebra such that $\mathcal{L}/\mathcal{V}$ does not commute with $\mathcal{V}$. This means that energy can be transferred to and from unobservable degrees of freedom. Once $\mathcal{V}$ has been calculated one can straightforwardly identify the dissipation channel as that part of the system Hamiltonian which lies in $i \mathcal{V}^{\perp}$ and does not commute with the whole. \\ \indent
Performing the observability decomposition on the Hamiltonian gives us $\hat H = \hat H_u + \hat H_o$, where $\hat H_u \in i \mathcal{V}^{\perp}$ and $\hat H_o \in i \mathcal{V}$. The total energy is then decomposed as
\begin{align}
E = \mathcal{O} + \mathcal{U}, \indent \mathcal{O} \triangleq \langle \hat H_o, \hat \rho \rangle , \indent \mathcal{U} \triangleq \langle \hat H_u, \hat \rho \rangle .
\end{align}
The main proposal of this paper is that differential heat and work flows can be identified as
\begin{align}
\dot Q \triangleq \frac{d \mathcal{U}}{dt}, \indent \dot W \triangleq \frac{d \mathcal{O}}{dt},
\end{align}
and that the component $\hat H_u \in i \mathcal{V}^{\perp}$ can be interpreted as a dissipation channel for the quantum control system under consideration provided that $[\hat H_u , \hat H_o] \neq 0$. \\ \indent Consider for a moment the heat flow when the Hamiltonian is time-independent,
\begin{align}
\dot Q = \langle \hat H_u, \frac{d}{dt} \hat \rho \rangle = -i \langle \hat H_u, [\hat H, \hat \rho] \rangle.
\end{align}
Employing a trace identity we obtain
\begin{align}
-i \langle [\hat H_u , \hat H], \hat \rho \rangle = -i \langle [\hat H_u , \hat H_o], \hat \rho \rangle.
\end{align}
If we define the operator
\begin{align}
\hat {\mathcal{D}} \triangleq i [\hat H_o , \hat H_u],
\end{align}
then $\langle \hat {\mathcal{D}}, \hat \rho \rangle = \dot Q$. In other words, the expectation value of $\hat {\mathcal{D}}$ is equal to the instantaneous transfer of energy from the observable to the unobservable component. This justifies interpreting it as a dissipation operator. \\ \indent
$\dot Q$ can also be calculated from quantities of $\mathcal{V}$, in case we don't know $\hat {\mathcal{D}}$. For a pure state we have
\begin{align}
\dot Q = - \sum_{i=1}^{\dim \mathcal{V}} \sum_{j=1}^{\dim \mathcal{V}} (\dot \theta_i h_j + \theta_i \dot h_j) F_{ij},
\end{align}
where we've introduced the notation
\begin{align}
\label{hk}
h_j (t) \triangleq \langle \hat H (t), \hat B_j \rangle,
\end{align}
and the $\theta_i$s and $F$ are defined as in Proposition \ref{fishprop}. If we define the matrix $\mathbb{H}_{ij} \triangleq \theta_i h_j$ we obtain
\begin{align}
\dot Q = - \langle \dot{\mathbb{H}} , F \rangle.
\end{align}
The right-hand side contains fluxes of observable quantities as well as a Fisher information, and the left is a dissipation measure. This opens up the possibility of using the Cramer-Rao inequality to obtain results similar to the Thermodynamic Uncertainty Relations discussed in the literature \cite{Horowitz2020ThermodynamicFluctuations}.
\begin{remark}
\label{rem}
The attentive reader will note that heat and work are traditionally viewed as path functions, while here they are functions of state. We believe this depends on whether the unobservable state component is explicitly modelled. If it is modelled, then the unobservable energy can be calculated at any initial and final times $t_f$ and $t_i$ to obtain the heat flow $Q = \mathcal{U} (t_f) - \mathcal{U}(t_i)$. However, if we are only modelling the observable part, then the transfer of energy between $i \mathcal{V}$ and $i \mathcal{V}^{\perp}$ have to be accounted for at each time step, making the heat flow an integral over the dissipation $Q = \int_{t_i}^{t_f} \langle \hat {\mathcal{D}} \rangle dt$.
\end{remark}

\begin{figure}[b]
\includegraphics[width=7cm]{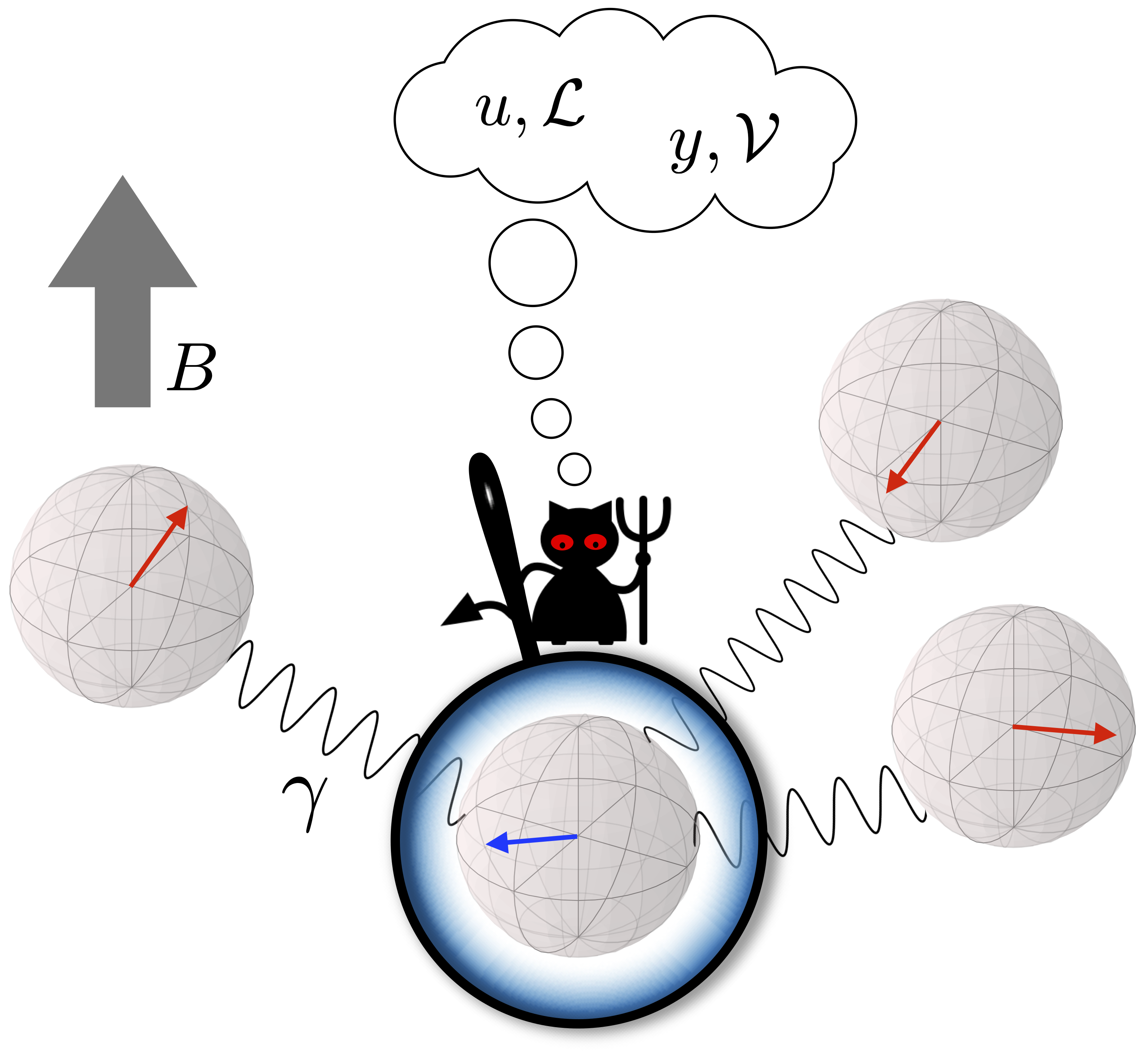}
\caption{\label{fig:deamon} An imperfect demon pondering his means for controlling and observing a central spin system. As full knowledge of the state is lacking, energy can leak out into unobservable degrees of freedom and thereby become ''degraded'', or ''entropic''.}
\end{figure}

\subsection{A Generalized Entropy}
Having defined a heat flow as a transfer of energy into $i \mathcal{V}^{\perp}$, we note that all such transfers are not qualitatively equal. Roughly speaking, we want to differentiate between heat flows associated with the state becoming ''more unobservable'', and those associated with a change in the occupied unobservable energy levels. This is essentially the difference between increasing the entropy and temperature respectively. We propose using the von-Neumann entropy $S_{vN}$ under a substitution to an output-equivalent state depending only on $\hat \rho_o$. This ''effective'' state is $\hat \rho_o + \frac{1}{n} \hat I$, and is the unique density matrix depending only on $\hat \rho_o$ which is indistinguishable from $\hat \rho$ in the sense of Definition \ref{defin}. Explicitly:
\begin{align}
\mathcal{S} (\hat \rho) \triangleq - \langle \hat \rho_o + \frac{1}{n} \hat I, \text{ log } (\hat \rho_o + \frac{1}{n} \hat I) \rangle.
\end{align}
It reduces to the ordinary von-Neumann entropy when $\mathcal{V} = su(n)$, and when $\mathcal{V} \neq su(n)$ it converges to the maximum value of $\log n$ as $\hat \rho_o \to 0$. The difference $\mathcal{S} (\hat \rho) - S_{vN} (\hat \rho)$ quantifies the information loss incurred in mapping to the effective state. \\ \indent
Define two operations $\mathfrak{O}$ and $\mathfrak{T}_t$ by their actions:
\begin{align}
\mathfrak{O}: \hat \rho \mapsto \frac{1}{n} \hat I - \sum_{i=1}^{\text{dim} \mathcal{V}}
\langle \hat \rho , \hat B_i \rangle \frac{\hat B_i}{\lvert \lvert \hat B_i \rvert \rvert^2}, \\
\mathfrak{T}_t : \hat \rho \mapsto \hat U_t \hat \rho \hat U_t^{\dag}.
\end{align}
They represent a projection onto the effective state (it is easy to verify that $\mathfrak{O}^2 = \mathfrak{O}$), and unitary time-evolution respectively. In general
\begin{align}
\mathfrak{T}_t (\mathfrak{O} (\hat \rho))  \neq \mathfrak{O} (\mathfrak{T}_t (\hat \rho)).
\end{align}
Since $\mathcal{S} = S_{vN} (\mathcal{O}(\hat \rho))$, and noting that the von-Neumann entropy is unitary invariant, it follows that if the commutator is zero then the time derivative of the entropy is zero as well. To see when this holds, set the commutator to zero and rearrange using the cyclic property of the trace to obtain
\begin{align}
\sum_{i=1}^{\text{dim} \mathcal{V}} \langle \hat \rho , \hat B_i \rangle \frac{\hat U_t \hat B_i \hat U_t^{\dag}}{ \lvert \lvert \hat B_i \rvert \rvert^2} = \sum_{i=1}^{\text{dim} \mathcal{V}} \langle \hat \rho , \hat U_t \hat B_i \hat U_t^{\dag} \rangle \frac{\hat B_i}{ \lvert \lvert \hat B_i \rvert \rvert^2}.
\end{align}
The above equation holds if it does so elementwise, i.e. if for every $i = 1,...,\text{dim} \mathcal{V}$ we have
\begin{align}
\langle \hat \rho , \hat B_i \rangle \mathfrak{T}_t (\hat B_i) = \langle \hat \rho , \mathfrak{T}_t (\hat B_i) \rangle \hat B_i.
\end{align}
This is true if $[\hat H, \hat B_i] = 0$ for all $\hat H \in i \mathcal{L}$ (which can be seen by taking the limit $t \to 0$), or if $\hat B_i$ is a simultaneous eigenvector for $\mathcal{L}$. If $\mathcal{V} \subset \mathcal{L}$, then the existence of such an eigenvector is guaranteed if $\mathcal{L}$ is solvable, by Lie's theorem. So a sufficient condition for the commutativity of $\mathfrak{O}$ and $\mathfrak{T}_t$ is that every basis vector of $\mathcal{V}$ either commutes with the whole of $\mathcal{L}$ or is a simultaneous eigenvector. \\ \indent
If we take the time-derivative of the entropy, assuming unitary dynamics,  we get
\begin{align}
\dot {\mathcal{S}} = - \langle \dot {\hat {\rho_o}}, \text{ log } (\hat \rho_o + \frac{1}{n} \hat I) \rangle = \\
\Bigl \langle \sum_{j=1}^{\text{dim} \mathcal{V}} \langle \hat \rho , [\hat H, \hat B_j] \rangle \frac{i \hat B_j}{\lvert \lvert \hat B_j \rvert \rvert^2}, \text{ log } (\hat \rho_o + \frac{1}{n} \hat I)) \Bigr \rangle.
\end{align}
We again see that $\dot {\mathcal{S}} \neq 0$ requires that at least some of the basis vectors of $\mathcal{V}$ are time-varying (i.e. lie outside the center of $\mathcal{L})$. We can unpack this equation further if we assume that $\mathcal{V}$ is an abelian, in which case $[\hat H , \hat B_j] = [\hat H_u , \hat B_j] $ for every $j$. Split the dissipation operator into a sum of dissipation operators, one for each basis vector of $\mathcal{V}$:
\begin{align}
\hat{\mathcal{D}} = \sum_{j=1}^{\text{dim} \mathcal{V}} \hat{\mathcal{D}}_j, \indent \hat{\mathcal{D}}_j \triangleq - i h_j \frac{[\hat H_u , \hat B_j]}{\lvert \lvert \hat B_j \rvert \rvert^2},
\end{align}
where $h_j$ is defined as in Equation \ref{hk}. Then we can write the entropy change as
\begin{align}
\label{clausius}
\dot {\mathcal{S}} = - \sum_{j=1}^{\text{dim} \mathcal{V}} \frac{\langle \hat{\mathcal{D}}_j \rangle}{h_j} \Bigr \langle \hat B_j , \text{ log } (\hat \rho_o + \frac{1}{n} \hat I)) \Bigr \rangle,
\end{align}
which clearly illuminates its relation to dissipation, in a manner reminiscent of Clausius' theorem.
\section{The Central Spin Model}
The central spin model consists of a distinguished spin $1/2$ particle, called the \emph{central spin}, connected to a bath of $N$ mutually non-interacting spin $1/2$ particles. The interaction is of Heisenberg XXX type with a ferromagnetic ground state. We have a product Hilbert space $\mathcal{H} = \mathcal{H}_2 \otimes \mathcal{H}_2 \otimes ... \otimes \mathcal{H}_2$ of $N+1$ factors, where the first factor represents the central spin, and $\text{dim} \mathcal{H} = 2^{N+1}$. The whole system is immersed in a constant magnetic field $B$ aligned to the $z$-axis and interacting equally with the central and bath spins, and the system is controlled by a tunable magnetic field along the $y$-axis acting on the central spin only. The Hamiltonian is a sum $\hat H (t) = \hat H_0 + \hat H_c (t)$ of drift and control terms
\begin{align}
\hat H_0 = - B \big ( \hat \sigma_z \otimes \hat I_b + \hat I_2 \otimes \sum_j \hat \sigma_z^{(j)} \big ) + \\
\sum_{j=1}^N \gamma_j \big (\hat \sigma_x \otimes \hat \sigma_x^{(j)} + \hat \sigma_y \otimes \hat \sigma_y^{(j)} + \hat \sigma_z \otimes \hat \sigma_z^{(j)} \big ), \\
\hat H_c (t) = - u(t) \hat \sigma_y \otimes \hat I_b,
\end{align}
where $\gamma_j < 0$ for all $j$, and the spin and identity operators on the bath are
\begin{align}
\hat \sigma_{i}^{(j)} \triangleq \underbrace{\hat I_2 \otimes ... \otimes \hat I_2}_\text{$j-1$ of them} \otimes \hat \sigma_i \otimes \underbrace{\hat I_2 \otimes ... \otimes \hat I_2}_\text{$N-j$ of them}, \\
\hat I_b = \underbrace{\hat I_2 \otimes ... \otimes \hat I_2}_\text{$N$ of them}
\end{align}
We are measuring the magnetization of the central spin in the $x$-direction: 
\begin{align}
\hat S = \hat \sigma_x \otimes \hat I_b.
\end{align}
Controllability, and therefore also observability, is strongly dependent on the relative values of the coupling parameters $\gamma_j$. If all of them are distinct the full system is both operator controllable and observable. We consider the opposite case where all of them are equal ($\gamma_j = \gamma_k$ $\forall j,k$) and the entire bath is behaving like a single collective spin. We refer the reader to \cite{Arenz2014ControlModel} for an extensive analysis of the controllability of the central spin model, and here just note that for equal coupling the system is uncontrollable and the dynamical Lie algebra satisfies
\begin{align*}
\text{dim} (\mathcal{L}) = \begin{cases} \frac{1}{6} (2+N)(9+4N(4+N)) \text{ for $N$ even} \\
\frac{1}{6} (1+N)(3+2N)(7+2N) \text{ for $N$ odd}
\end{cases}
\end{align*}
The central spin, however, is locally controllable, which together with $\hat S$ being a local observable, implies that $\mathcal{V} \subset \mathcal{L}$ is an ideal.

\subsection{Numerical analysis}
\begin{figure}[ht!]
\includegraphics[width=9cm]{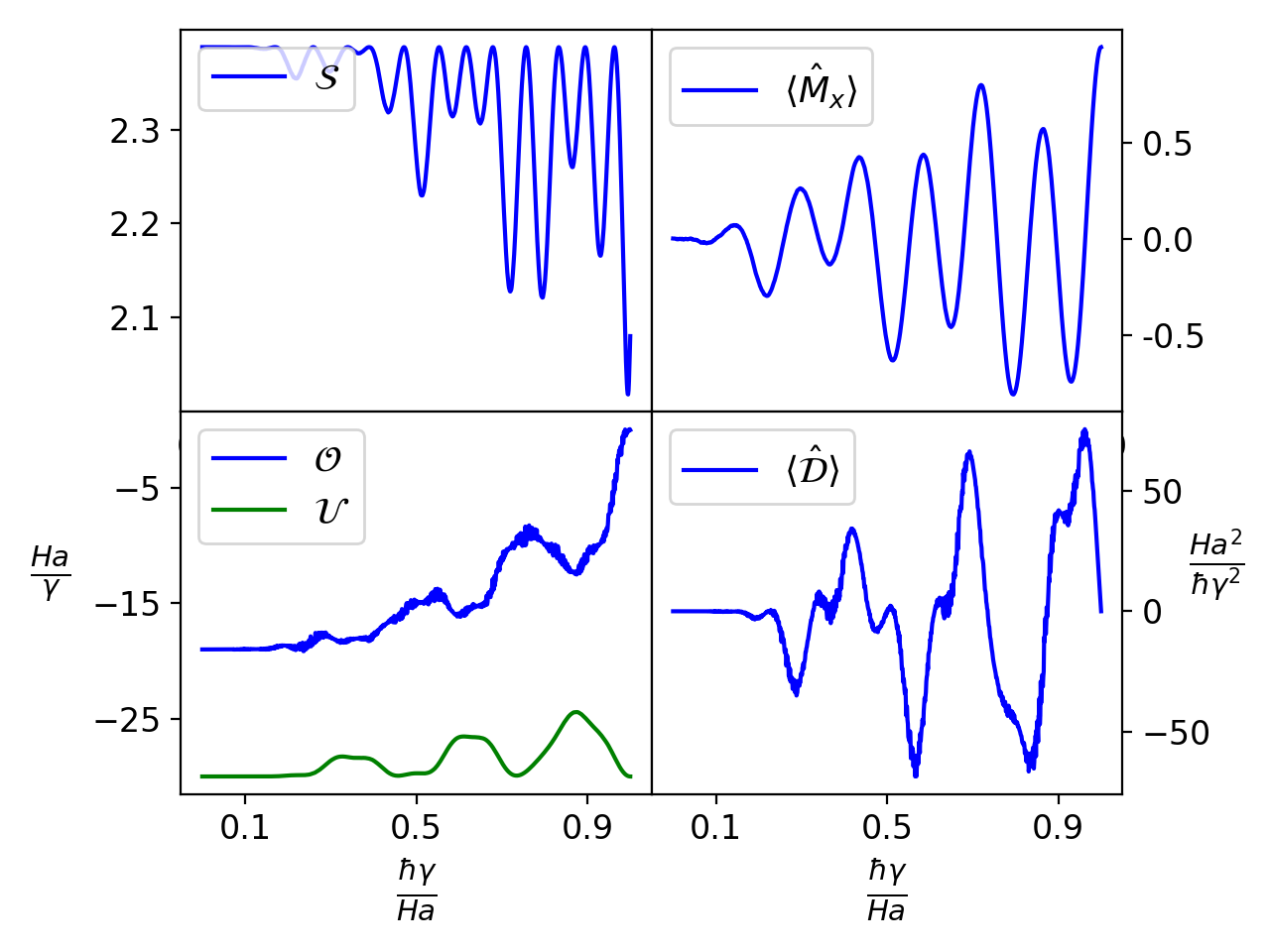}
\caption{\label{fig:controlled} Evolution of the central spin model during the controlled phase. Top left: generalized entropy $\mathcal{S}$. Top right: magnetization $\langle \hat M_x \rangle$ along the $x$-axis for the central spin. Bottom left: observable and unobservable energies $\mathcal{O}$ and $\mathcal{U}$. Bottom right: instantaneous dissipation $\langle \hat {\mathcal{D}} \rangle$.}
\end{figure}

We let the number of bath spins ($N$) be equal to $3$, giving us a Hilbert space dimension of $2^4 = 16$. We consider an initial state with all spins perfectly aligned to the constant magnetic field along the $z$ direction. This is a pure state, and it will remain pure throughout the simulation. A control field, here a variable magnetic field along the $y$ direction applied to the central spin, is obtained using the pulse-optimizer function available in Qutip \cite{Johansson2013QuTiPSystems,Johansson2012QuTiP:Systems} to drive the central spin towards alignment with the $x$ axis at terminal time $\tau$. In other words, this is an optimal control problem of Mayer type with target functional
\begin{align}
J [u] = \langle \hat S , \hat \rho (\tau) \rangle.
\end{align}
Afterwards the system evolves freely under a Gaussian control field for another time period of $\tau$. During both processes we compute the observability decomposition of the energy $(\mathcal{U},\mathcal{O})$, the instantaneous dissipation $\langle \hat {\mathcal{D}} \rangle$, the entropy $\mathcal{S}$ and the measured output $\langle \hat S \rangle$. \\ \indent
We are working in Hartree atomic units, and moreover, calculate the energy in units of $\frac{\text{Ha}}{\gamma}$, giving time units of $\frac{\hbar \gamma}{\text{Ha}}$. In these units, we set the magnetic field $B$ equal to $10$ and the interaction constant $\gamma$ equal to $-3$. The terminal time $\tau$ is set to unity, and the interval is discretized into $1000$ timeslots for the optimization procedure and time-evolution. During the free evolution phase the control field is set equal to a Gaussian centered at zero with a standard deviation of $0.1$. \\ \indent
The dynamical Lie algebra and the observability space are calculated numerically using the algorithm presented in Appendix B. We find that $\text{dim} (\mathcal{L}) = 78$, in agreement with the results in \cite{Arenz2014ControlModel}, and has a maximum depth of $11$. For the observability space we find $\text{dim} (\mathcal{V}) = 41$ with a maximal depth of $1$. The basis of the observability space is orthogonalized using the Gram-Schmidt procedure. \\ \indent
Figure \ref{fig:controlled} shows the evolution of the controlled system in the interval $[0, \tau]$. The heat, work and entropy changes are:
\begin{align}
Q_c = 0.00, \indent W_c = 19.00, \indent \Delta \mathcal{S}_c = - 0.29.
\end{align}

During the controlled phase the entropy begins at its maximal value, which is to be expected since the system is initialized in a state with zero output. It oscillates with the evolution of the system, attaining lower values the larger the absolute value of the output. Initially there is no dissipation of energy, but $\langle \hat{\mathcal{D}} \rangle$ increases with the amplitude of oscillation of the observable output as the system is controlled towards the desired state. Overall, $\mathcal{O}$ increases substantially during, while $\mathcal{U}$ is almost constant. \\ \indent
\begin{figure}[ht!]
\includegraphics[width=9cm]{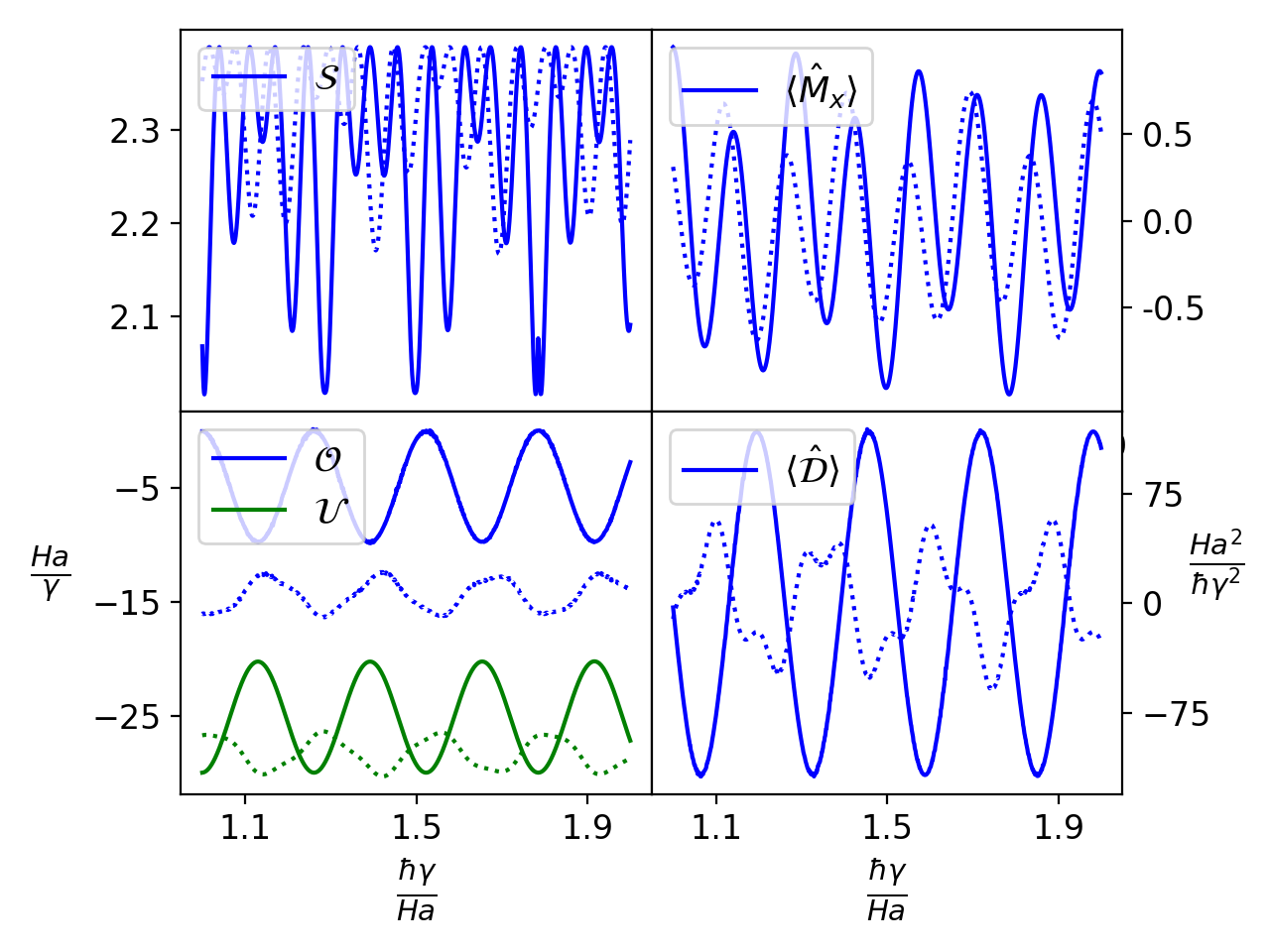}
\caption{\label{fig:freeevo} Evolution of the central spin model during the free evolution phase. The quantities shown are the same as in Figure \ref{fig:controlled}. For comparison, the dotted lines show a system that evolves freely from the state obtained at $t = 0.6$ during the controlled phase.}
\end{figure}
Figure \ref{fig:freeevo} shows the subsequent free evolution in the interval $[\tau , 2 \tau]$, during which energy flows sinusoidially between the two subspaces. These oscillations may be interpreted as thermal fluctuations. This behavior is expected as isolated quantum systems undergo periodic motion. It also shows the free evolution of the state which is obtained at $t= 0.6$ during the controlled phase. The purpose is to illustrate that when we control the system to reduce the entropy, the more we reduce the entropy the greater the amplitude of the oscillations afterwards. The amplitudes of oscillations of the other quantities are similarly affected. In conclusion, with external control we can increase the size of thermal fluctuations.

\newpage
\section{Conclusion}
We have proposed a mathematically explicit method for distinguishing heat from work dependent on the sensor and control capabilities at hand, based on a control-theoretic observability decomposition due to D. D'Alessandro. We also derived a dissipation operator, and proposed an accompanying entropy function which for fully controllable and observable systems reduces to the ordinary von-Neumann entropy, but which can also yield non-zero values for pure states if the observables under consideration are informationally incomplete (i.e. if the state estimation problem is not solvable). And perhaps more importantly, this entropy can change also for an isolated system. \\ \indent
Thermodynamics is fundamentally the science of the difference between heat and work, and the possibilities of their interconversion. This is the key distinction that the whole edifice is built around, so a thorough understanding of it is of paramount importance for the development of the entire field. A proper understanding of the heat-work decomposition will allow us to extend thermodynamics further into the quantum realm, improve our understanding of dissipation and thermalization, and shed light on some of the largest conceptual problems in physics. \\ \indent
The nature of the heat-work decomposition is a major open problem in quantum thermodynamics whose solution would have fundamental ramifications for both pure and applied science. We believe that our proposal provides the correct answer to the conundrum.
\begin{acknowledgments}
We acknowledge funding from the Swedish Research Council (VR) grant 2019-03901, and the European Research Council (ERC) through the Synergy Grant 854843 - FASTCORR. We thank Yishao Zhou (Stockholm University) for being an excellent teacher and for encouraging the lead author to study quantum control and pursue this project. Erik Sjöqvist is acknowledged for fruitful  discussions. 
\end{acknowledgments}


\appendix
\section{}
\subsection{Notation $\&$ conventions}
\begin{itemize}
    \item The inner-product is always of Hilbert-Schmidt type: $\langle A,B \rangle = \text{Tr} (A^{\dag} B)$.
    \item We work in Hartree atomic units where $\hbar = e = m_e = k_e = 1$.
    \item The anti-commutator is denoted $[A,B]_+ = AB + BA$.
    \item The Lie algebra $\textbf{u}(n)$ is the set of skew-Hermitian matrices of dimension $n$, and $\textbf{su}(n)$ is the subset of those which are traceless. The corresponding Lie groups are $U(n)$, the set of unitary matrices of the same dimension $U(n)$, and the subset of those with determinant unity, $SU(n)$.
    \item $\hat I_2$ is the two-dimensional identity matrix.
    \item $\text{ad}_A (B)$ is the set of all commutators $[a,b]$ where $a \in A$ and $b \in B$. If there is a superscript $\text{ad}^i$ it represents a set formed by taking $i$ repeated commutators of elements in $B$ with elements in $A$. For $i=2$ the elements are of the form $[a_2,[a_1,b]]$.
\end{itemize}
\subsection{Algorithm for computing $\mathcal{L}$ and $\mathcal{V}$}
For a bilinear Hamiltonian of the form
\begin{align}
\hat H (t) = \hat H_0 + \sum_{j=1}^k u_j (t) \hat H_{c,j},
\end{align}
the dynamical Lie algebra $\mathcal{L}$ is defined as the Lie algebra generated by the set $\{ i \hat H_1, i \hat H_{c,1}, ... , i \hat H_{c,k} \}$. We denote the generating set by $G$. Denote the set of basis elmements of our would be algebra $\mathcal{B}_{\mathcal{L}}$. The first step in the algorithm is to
\begin{itemize}
\item \textbf{Step 0: }Add all linearly independent elements of $G$ to $\mathcal{B}_{\mathcal{L}}$. These are the elements $\mathcal{L}$ of depth $0$.
\item \textbf{Step k: } Compute the set of all commutators of $B_{\mathcal{L}}$ with $G$. Each result which is linearly independent of the rest of $\mathcal{B}_{\mathcal{L}}$, add it to $\mathcal{B}_{\mathcal{L}}$. These are the basis elements of depth $k$. If no new linearly independent elements are obtained, the algorithm terminates and $\mathcal{B}_{\mathcal{L}}$ is a basis of $\mathcal{L}$. Otherwise, proceed to step $k+1$.
\end{itemize}
The algorithm to compute a basis $\mathcal{B}_{\mathcal{V}}$ of $\mathcal{V}$ is identical to the above, except that the generating set is $\mathcal{L}$ and the depth $0$ element(s) the measured observable(s). In other words, add your measured observable to $\mathcal{B}_{\mathcal{V}}$. Then compute all commutators of depth $1$ of $\mathcal{L}$ with $\mathcal{B}_{\mathcal{V}}$, and add the linearly independent results to $\mathcal{B}_{\mathcal{V}}$. Then just repeat the procedure until no new linearly independent elements are generated.
\subsection{Proof of Proposition 1}
Assume that $\text{dim}(\mathcal{V}) = r$ and let $\rho$ be written in the form
\begin{align*}
\hat \rho (\theta) = \hat \rho_u + \sum_{j=1}^r \theta_j \hat B_j,
\end{align*}
where we've introduced the notation $\hat B_j$ for the basis elements of $i \mathcal{V}$, and $\theta_j \triangleq \text{Tr} \{ \hat B_j \hat \rho \}$. The $r$-parameter estimation problem is now formulated as a problem of finding the projections of $\hat \rho$ along every basis operator of $i \mathcal{V}$. Taking the partial derivative of $\hat \rho$ with respect to one of the parameters we obtain
\begin{align*}
\frac{\partial}{\partial \theta_j} \hat \rho (\theta) = \hat B_j = \frac{1}{2} [\hat L_{\theta_j} , \hat \rho]_{+} = \frac{1}{2} (\hat L_{\theta_j} \hat \rho + \hat \rho \hat L_{\theta_j}).
\end{align*}
Since the basis operators $\hat B_j$ are traceless, we see that the expectation values of the SLD operators vanish, i.e.
\begin{align*}
\langle \hat L_j \rangle = \text {Tr} \{ \hat L_{\theta_j} \hat \rho \} = \text{Tr} \{ \hat B_j \} = 0.
\end{align*}
By employing the cyclic property of the trace, a straightforward calculation shows that
\begin{align*}
\text{Tr} \{ [\hat B_i , \hat B_j]_+ \} = \text{Tr} \{ 2 \hat L_{\theta_i} \hat \rho \hat L_{\theta_j} \hat \rho + \hat \rho^2 \hat L_{\theta_i} \hat L_{\theta_j} + \hat \rho^2 \hat L_{\theta_j} \hat L_{\theta_i}  \}.
\end{align*}
If the quantum state is pure, then the density matrix is a projector with $\hat \rho^2 = \hat \rho = \lvert \psi \rangle \langle \psi \rvert$, and
\begin{align*}
 \text{Tr} \{ 2 \hat L_{\theta_i} \hat \rho \hat L_{\theta_j} \hat \rho \} = 
 2 \text{Tr} \{ \hat L_{\theta_i} \lvert \psi \rangle \underbrace{\langle \psi \rvert \hat L_{\theta_j} \lvert \psi \rangle}_{\text{constant}} \langle \psi \rvert \} = \\
 2 \langle \psi \rvert \hat L_{\theta_j} \lvert \psi \rangle \text{Tr} \{ \hat L_{\theta_i} \lvert \psi \rangle \langle \psi \rvert \} = 
 2 \langle \hat L_{\theta_j} \rangle \langle \hat L_{\theta_i} \rangle = 0.
\end{align*}
Writing out the Fisher information explicitly
\begin{align*}
F_{i , j} = \frac{1}{2} \text{Tr} \{ \hat \rho \hat L_{\theta_i} \hat L_{\theta_j} + \hat \rho \hat L_{\theta_j} \hat L_{\theta_i} \},
\end{align*}
we now see that under the assumption of purity,
\begin{align*}
F_{\theta_i , \theta_j} = \frac{1}{2} \text{Tr} \{ [\hat B_i , \hat B_j]_+ \}.
\end{align*}
By the cyclic property of the trace, and the hermiticity of the matrices $\{ \hat B_i \}$, this is equal to $\langle \hat B_i , \hat B_j \rangle$. $\square$


%

\end{document}